\documentclass[epj]{svjour}
\usepackage{graphics}
\usepackage{dcolumn}
\usepackage{bm}
\usepackage{multirow}
\usepackage{array}
\usepackage{pbox}
\usepackage{booktabs}
\usepackage[numbers]{natbib}
\usepackage{mathtools}
\usepackage{url}
\begin{document}
\title{Pileup mitigation at the Large Hadron Collider \\ with Graph Neural Networks}

\author{J.~Arjona Mart\'inez\inst{1} 
  \and O.~Cerri\inst{2} \and M.~Spiropulu\inst{2} \and JR.~Vlimant\inst{2} \and M.~Pierini\inst{3}}

  
\institute{University of Cambridge, Trinity Ln, Cambridge CB2 1TN, UK \and California Institute of Technology, 1200 E.\ California Blvd, Pasadena, CA 91125 \and CERN, CH-1211 Geneva, Switzerland \\ \email{maurizio.pierini@cern.ch}}%

%
%
\abstract{
At the Large Hadron Collider, the high transverse-momentum events  studied by  experimental collaborations occur in coincidence with parasitic low transverse-momentum collisions, usually referred to as pileup. Pileup mitigation is a key ingredient of the online and offline event reconstruction as pileup affects the reconstruction accuracy of many physics observables. We present a classifier based on Graph Neural Networks, trained to retain particles coming from high-transverse-momentum collisions, while rejecting those coming from pileup collisions. This model is designed as a refinement of the {\tt PUPPI} algorithm~\cite{PUPPI}, employed in many LHC data analyses since 2015. Thanks to an extended basis of input information and the learning capabilities of the considered network architecture, we show an improvement in pileup-rejection performances with respect to state-of-the-art solutions.
} 
\maketitle

\section{Introduction}
\label{sec:intro}

In order to deliver large datasets to experimental collaborations, the CERN Large Hadron Collider (LHC) operates with proton bunches, each containing as many as ${\cal O}(10^{11})$ protons. These protons are densely packed to increase the beam luminosity and, with it, the collision rate. This luminosity increase comes at the cost of an increasing number of parasitic collisions (pileup), typically consisting of soft-QCD events at small transverse momentum ($p_T$). At the end of LHC Run II, an average of $\sim 40$ (and a maximum of $\sim 80$) collisions happened simultaneously to each interesting high-$p_T$ event.

Due to the shape of the luminous region, the $(x,y)$ coordinates\footnote{We use a Cartesian coordinate system with the $z$ axis oriented along the beam axis, the $x$ on the horizontal plane, and the $y$ axis oriented upward. The $x$ and $y$ axes define the transverse plane, while the $z$ axis identifies the longitudinal direction. The azimuth angle $\phi$ is computed with respect to the $x$ axis. The polar angle $\theta$ is used to compute the pseudorapidity $\eta = -\log(\tan(\theta/2))$. We fix units such that $c=\hbar=1$.} of the pileup collisions in an event are aligned to those of the high-$p_T$ interesting collision, referred to as the leading vertex (LV). On the other hand, pileup vertices can be displaced by several cm from the LV in the $z$ direction. The impact of pileup collisions is reduced by applying targeted algorithms designed to estimate and mitigate this effect.

\begin{figure}[tb]
\centering 
\includegraphics[width=.95\textwidth]{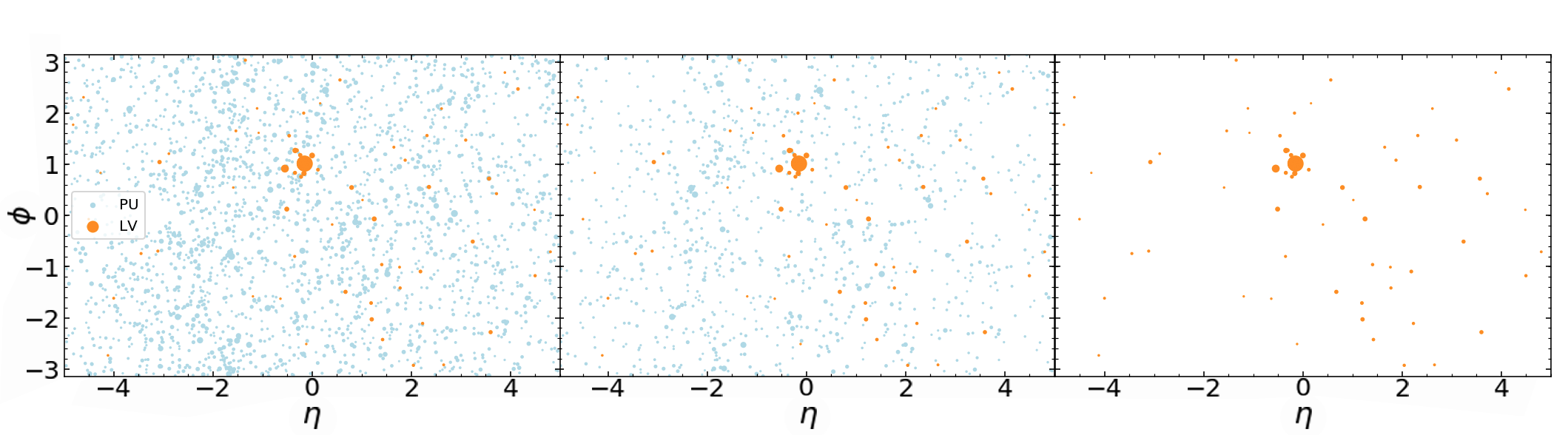}
\caption{Depiction of the effect of CHS. The full event (left), the event after CHS is applied (middle) and the Ground Truth (right) are shown. A $Z\to\nu\overline{\nu}$+jets event is superimposed to 80 pileup events. Particles from the LV are shown in orange (dark) and those from pileup in blue (light).}
\label{fig:CHSintro} 
\end{figure}

Being detected in the inner layers of a typical multi-purpose detector such as ATLAS~\cite{Aad:2008zzm} or CMS~\cite{Chatrchyan:2008aa}, charged particles can be tracked back to their point of origin, while this is not typically the case for neutral particles. Thanks to the CMS~\cite{CMSCHS} and ATLAS~\cite{ATLASCHS} vertex resolution, charged particles from pileup can then be associated to vertices other than the LV and consequently removed. This technique, referred to as charged hadron subtraction (CHS), greatly simplifies the problem, as can be seen in Figure~\ref{fig:CHSintro}. The main challenge becomes correcting for the neutral pileup contribution, for which sufficient vertex information is typically unavailable. Early approaches, such as the area-subtraction method~\cite{AreaMethod2,AreaMethod3,AreaMethod4,AreaMethod1} employed in LHC Run I (2009-2012)  analyses, correct the event based only on the characteristic per-event pileup energy density. While these approaches help in obtaining unbiased estimates of the jets four-momenta, they are affected by a serious resolution loss for a large number of pileup interactions, even when extended to jet shapes~\cite{areasubtractionjetshapes}. This motivated the introduction of new algorithms for the LHC Run II (2015-2018). 

The currently adopted pileup mitigation techniques consist of rule-based algorithms and usually operate on a per-particle basis, tailored to suppress particles believed to originate from pileup interactions, or to weight them proportionally to their probability of originating from the hard interactions. Examples of the former category include the {\it SoftKiller} algorithm~\cite{Softkiller}. The PileUp Per Particle Identification ({\tt PUPPI}) algorithm~\cite{PUPPI} employed by the CMS collaboration in the LHC Run II is an example of the latter category. These two algorithms 
fairly represent the state-of-the-art performance for what concerns pileup mitigation algorithms, having being successfully tested on real LHC collision data~\cite{CMS-PAS-JME-14-001,ATLAS-CONF-2017-065}. For this reason, we take them as a baseline in this paper. 

Our study takes as starting point the {\tt PUPPI} algorithm. Unlike many other pileup-removal algorithms, {\tt PUPPI} is designed to assign a weight to each particle, the so-called {\tt PUPPI} weights. These weights quantify how likely it is that each particle originated from the LV. As described in Ref.~\cite{PUPPI}, the weight computation is based on the per-event distribution of the quantity:
\begin{equation}
\alpha_i^\beta = \log \sum_{j \in \textrm{event}} \xi_{ij} \times \Theta(\Delta R_{ij} < R_{\textrm{min}}) \times \Theta(\Delta R_{ij} < R_0)~.
\label{eq:alpha}
\end{equation}
In Eq.(\ref{eq:alpha}), $i$ is the label of the considered particle in the event and $\xi_{ij}=p_{Tj}/\Delta R_{ij}^\beta$; $\Delta R_{ij} = \sqrt{\Delta \phi^2 + \Delta \eta^2}$ is the distance between the $i$-th and $j$-th particle in the plane identified by the pseudorapidity $\eta$ and the azimuth angle $\phi$; $R_0=0.3$ defines a cone around the $i$-th particle, and $R_{\textrm{min}}=0.02$ removes the region surrounding the $i$-th particle. In Ref.~\cite{PUPPI}, $\alpha^1$ is found to be the optimal metric to quantify the {\tt PUPPI} weight. When CHS is applied upstream to {\tt PUPPI}, the sum in Eq.(\ref{eq:alpha}) is performed over the charged particles from the LV, as opposed to the full event. By definition, the {\tt PUPPI} algorithm requires to be tuned on the specific dataset it is applied to. For a fair comparison, the {\tt PUPPI} results presented in this work are obtained applying the tuning procedure outlined in Ref.~\cite{PUPPI} to the datasets described in Section~\ref{sec:data}.

Our work aims to extend the traditional {\tt PUPPI} setup using Deep Learning. In particular, we present a classifier, based on Graph Networks, trained to identify particles originating from the high-$p_T$ events and discard the others. We consider an architecture based on Gated Graph Neural Networks (GGNN)~\cite{GGNN} as our final model, called {\tt PUPPIML}. In addition, a model based on fully-connected Neural Networks~\cite{MLP} and one based on Gated Recurrent Units (GRU)~\cite{GRU} are presented for comparison.

This paper is structured as follows: Section~\ref{sec:related_work} describes related literature. Section~\ref{sec:data}~and~\ref{sec:models} describe the utilized data and models, respectively. Results are presented in Section~\ref{sec:results}, while the robustness of the proposed approach is discussed in Section~\ref{sec:robustness}. Conclusions are given in Section~\ref{sec:conclusions}.

\section{Related work}
\label{sec:related_work}

Machine Learning (ML) traditionally played a prominent role in High Energy Physics (HEP), as discussed for instance in Ref.~\cite{naturereview}. Among the many existing proof-of-principle studies, few applications have already been deployed in the central data processing of major HEP experiments. For instance, recurrent architectures proved fruitful in bottom-jet identification~\cite{CMS-DP-2017-027, ATL-PHYS-PUB-2017-003} and Convolutional Neural Networks (CNNs) in neutrino physics~\cite{Aurisano:2016jvx}. Graph Networks have very recently been used for jet tagging, matching the performances of other deep learning approaches~\cite{graphjettagging}, and to identify interesting typologies at the LHC~\cite{Abdughani:2018wrw}, and in IceCube~\cite{Choma2018GraphNN}.

A first application of machine learning to the problem of pileup subtraction is presented in Ref.~\cite{PUMML}, describing the PUMML algorithm. This work clearly demonstrates that the use of Deep Learning techniques to pileup removal results in a performance improvement. The PUMML algorithm is based on CNNs. To apply it, one needs to represent the collision event as an image. This is usually done binning the portion of the ($\eta$, $\phi$) plane covered by detector acceptance and creating a map of the deposited $p_T$ in each bin. Such a practice poses some problem when confronting the main features of a typical collider detector. For instance::
\begin{enumerate}
\item PUMML  is applied to images of cropped regions around the reconstructed jets, derived by binning the ($\eta$, $\phi$) plane. The binning choice is intended to be ``representative of typical tracking and calorimeter resolutions''~\cite{PUMML}. This approach is very suitable for the central barrel of a typical detector, but it overlooks the complexity of a typical detector geometry. In particular, it neglects the existence of complicated overlap regions between the barrel and the endcap regions, and the irregular geometry of the endcaps (e.g., the variable segmentation of the CMS ECAL and HCAL endcaps). On the other hand, pileup subtraction is particularly important in the endcap regions, where the solenoid magnetic field of a typical cylindrical detector pushes the abundant low-$p_T$ particles produced in pileup interactions. 
\item The use of a calorimeters-inspired  ($\eta$, $\phi$) fixed-size grid neglects the fact that charged particles are mainly detected through the inner tracker. A typical-tracker ($\eta$, $\phi$) resolution can hardly be represented as a fixed number, since it depends on the track $p_T$ (as shown for instance in Ref.~\cite{Chatrchyan:2014fea} for the CMS detector). This fact might not be crucial for a “calorimeter-centric” jet reconstruction. On the other hand, such an assumption is extremely sub-optimal for local or global reconstruction algorithms based on particle flow~\cite{Aaboud:2017aca,CMS_PF}. In these cases, the enhanced tracker resolution plays a crucial role in driving the measurement of particle momenta. 
\item The assumed ($\eta$, $\phi$) resolution (0.025x0.025) is realistic for the barrel region of a typical electromagnetic calorimeter, but substantially underestimates a realistic tracker ($\eta$, $\phi$) resolution. As an example, the ($\eta$, $\phi$) resolution of the CMS tracker for a 1~GeV (10~GeV) track is $\sim 10$ ($\sim 100$) times better than what assumed in Ref.~\cite{PUMML}. Even if a fixed-size regular ($\eta$, $\phi$) grid could faithfully represent the ($\eta$, $\phi$) resolution of a silicon-based tracker device, one would be confronted with a very large number of cells and, consequently, many technical complications in training the algorithm (e.g., GPU memory consumption and image sparsity).
\end{enumerate}
While these might be marginal problems for traditional approaches to event reconstruction, we see them as extremely sub-optimal for the CMS global event description, which is what we are more familiar with. This fact motivated us to extend the work of Ref.~\cite{PUMML}, investigating network architectures that would not rely on an image-based representation of the event or, more generally, on specific assumptions on the detector geometry or granularity. We consider three different network architectures: a fully-connected deep neural network (DNN), a network with GRU layers~\cite{GRU}, and a GGNN~\cite{GGNN}. By utilizing feature- or particle-based representations of the event, these three architectures don't rely on specific aspects of the underlying detector geometry and can be easily integrated in a global or local particle-flow reconstruction like those employed at the LHC~\cite{Aaboud:2017aca,CMS_PF}.


\section{Dataset}
\label{sec:data}

The dataset employed in this work consists of simulations of LHC proton-proton collisions at a center-of-mass energy of 13~TeV, created using the {\tt PYTHIA 8.223}~\cite{pythia} event generator, tune 4C~\cite{Corke:2010yf}. The generated events correspond to a sample of $Z$ bosons decaying to a pair of neutrinos and produced in association with at least one quark or gluon, resulting in at least one jet. Following Refs.~\cite{PUPPI,Softkiller}, the generation of the underlying event is turned off, so that the reconstructed particles in the event could be divided into two well defined categories: particles from the high-$p_T$ LV or pileupparticles. This technical aspect facilitates the sample preparation and training while having a negligible impact on performances. The list of particles originating from the hard collision and the consequent shower of quarks and gluons are given as input to the {\tt DELPHES 3.3.2}~\cite{delphes} detector-simulation software. As typically done in previous studies of this kind~\cite{PUPPI,Softkiller,PUMML}, no detector reconstruction effect is applied at this stage. {\tt DELPHES} is mainly used as a convenient tool to read the {\tt HEPMC} files generated by {\tt PYTHIA}, overlay the pileup events, and store the event content in a {\tt ROOT}~\cite{root} file, preserving the provenance information for each particle (LV or pileup events). We set the average number of pileup interactions $\overline{n}_{\textrm{PU}}$ to 20, 80 or 140, and randomly generate the per-event pileup multiplicity according to a Poisson distribution. A corresponding number of events is then sampled from a so-called minimum-bias library, consisting of an inclusive sample of {\tt PYTHIA}-generated QCD events. The $Z$+jets events are generated in batches of 100 events each. For each batch, pileup events are samples from a batch-specific library of 10000 events. This allows to reduce the probability of event re-usage to a negligible level, preventing the network to learn specific patterns of specific events in the training datasets. 

All final-state particles except neutrinos are kept and assumed to be massless, as normally done for high-$p_T$ physics studies. For each particle, the following information is stored:
\begin{itemize}
\item The momentum coordinates ($p_T$, $\eta$, $\phi$).
\item The electric charge of the particle.
\item The local shape $\alpha_i^\beta$ for $\beta = 1, 2, 3$, as defined in Ref.~\cite{PUPPI} (see Eq.~\ref{eq:alpha}),  both summing over the full event ($\alpha_i^{\beta,F}$) or only over the charged particles from the LV ($\alpha_i^{\beta,C}$).
\item The {\tt PUPPI} weight associated to the particle: a number between 0 and 1 that can be related to the probability of the particle being pileup.
\item A flag set to 1 for charged particles from the LV, -1 for charged pileup particles and 0 for all neutral particles. This flag provides a simple encoding for when CHS is used.
\item A {\it pileup flag} indicating whether the particle belongs to the hard scattering or to any of the pileup vertices. This information is used as the ground truth later on. 
\end{itemize}
We assume units such that $\hbar = c = 1$. Furthermore, we store for each event the 
median $p_T$ per unit area in $(\eta,\phi)$ for all particles ($\rho$), for all charged particles alone ($\rho_C$), and for all neutral particles alone ($\rho_N$).

\begin{figure}[tb]
\centering 
\includegraphics[width=.9\textwidth]{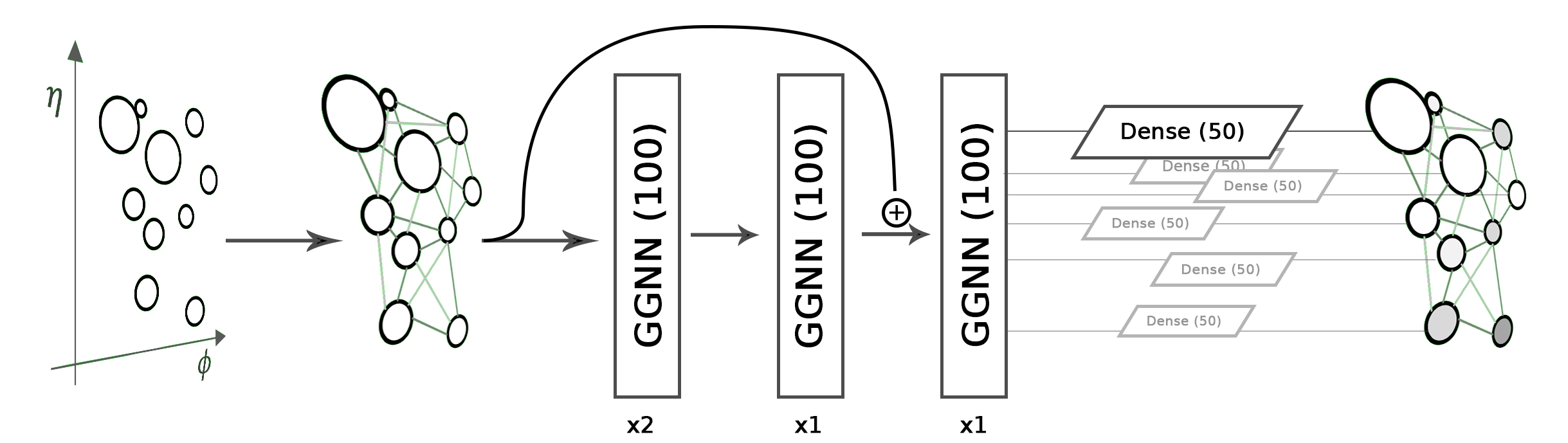}
\hfill
\caption{\label{fig:model} Conceptual depiction of the GGNN model architecture. The event is pre-processed by linking local particles together, after which it is fed to 3 GGNN layers with time-steps [2, 1, 1] and including a residual connection from the first to the third layer. This is then passed, individually per graph node, to a fully-connected network that outputs a [0,1] pileup classification score. Adam is used with a learning rate of 0.004 to minimize the binary cross-entropy. The output of the network is checked to be a well-calibrated probability.}
\end{figure}

\section{Network architectures}
\label{sec:models}

{\tt PUPPI} can be straightforwardly interpreted as a per particle classification algorithm. Under this point of view, traditional metrics such as the Receiver Operating Characteristic (ROC) curve (true positive rate against false positive rate) or the accuracy (fraction of correctly classified particles) can be used. The choice of the shape variable $\alpha^1$ is then driven by its discriminating power, with the underlying assumption that a better classification performance should correlate with a better reconstruction of physics-motivated quantities which are relevant to study these data. For all the investigated network architectures, we generalize this approach to multiple shape variables, indicated from now on as features. We feed as input to our networks all the particles, with all the features discussed in Sec.~\ref{sec:data} except for the pileup flag, which we use as the training ground truth. The global features are concatenated to each particle's individual features. An generalization of {\tt PUPPI} by mean of ML techniques is already discussed in Ref.~\cite{PUPPI}, where it is asserted that training a Boosted Decision Tree modestly improves performance when compared to the use of $\alpha^1$ as discriminating quantity.

Our most straightforward model makes use of two stacked fully-connected hidden layers and a final single-neuron layer with a sigmoid activation function. This network is trained, as all the other models, to minimize a binary cross-entropy loss function using the Adam optimizer~\cite{Adam}. This model stands out for its simplicity, as it operates on each particle completely independently of the others, but suffers from a clear issue: while the input includes global ($\rho, \rho_C, \rho_N$) and local ($\alpha_i^C, \alpha_i^F$) features, the network has no mechanism by which it could learn these or similar features.

Extending the network architecture beyond a simple per-particle processing, one could overcome this limitation. To this purpose, different network architectures can be chosen. Reference~\cite{PUMML} describes an approach based on CNNs. Motivated by the arguments described in Sec.~\ref{sec:related_work}, we complement the results of Ref.~\cite{PUMML} by studying GRUs and GGNNs. Both these architectures take as input the full list of particles in the event, outputting a per-particle label.

The GRU is a recurrent neural network architecture that sequentially processes each item of an input list, based on the outcome of previous-item processing.  While making no assumption on the underlying detector geometry, the GRU architecture implies the use of a ranking principle to order the items in the input list. In our study, the inout list contains the particles in the event, which are ordered by their $p_T$ value. This is one of the many arbitrary choices that one could make. In the network, we make use of a bidirectional GRU layer, i.e., we consider both increasing- and decreasing-$p_T$ ordering. The output of this layer is concatenated to each particle's features. We show that this approach does not improve the classification performance with respect to DNNs and traditional methods. This is mainly due by the fact that GRUs require a global ordering criterion, while the information determining if a given particles belongs to the LV or originates from pileuphas mainly to due with the particle's local neighborhood. 


The architecture of a GGNN (and in general any graph network)  provides the capability of learning specific features of an input item from the nearby neighbors, delivering state-of-the-art performance (see Section~\ref{sec:results}) without making any assumption on the underlying geometry or input ordering. A GGNN can be considered a special type of Message Passing Neural Network~\cite{MessagePassing}. Each example in the input dataset is represented as a set of vertices $v \in \mathcal{V}$. The vertices are connected by directional edges $e_t \in \mathcal{E}$. The combination of the vertices and the connection edges forms the graph $\mathcal{G} = (\mathcal{V}, \mathcal{E})$. At input, each vertex of the graph is represented by a feature vector $\mathbf{h}_v^0$ of dimension $h^0$. During a graph-layer processing, the graph evolves in a set of time steps. Each time step $i>0$ consists of two operations:
\begin{enumerate}
    \item For each node $v$, we generate an incoming message from each connected neighbour vertex $v_j$. The message $\mathbf{a}_{v,v_j}^i$ is computed multiplying a learnable $h^{i-i} \times h^{i-i}$ matrix $\mathbf{A}_t$ by the feature vector $v_j$ from the previous time interaction:
      \begin{equation}
      \label{eq:ggnn:2}
      \mathbf{a}_{v,v_j}^i = \mathbf{A}_t \mathbf{h}_{v_j}^{i-1}, \forall v_j : v_j \xrightarrow{\text{t}} v~.
      \end{equation}
    When different kinds of edges are present in the graph, edge-type specific A matrices are used.
    \item All messages coming to a node $v$ are averaged. The average is used to create a new representation of the node. Repeating this operation for each node, a new graph is generated, isomorphic to the original one but characterized by a new node representation, learned through the message exchange. In the specific case of a GGNN, the node update is performed utilizing one step of a GRU, initializing the GRU memory cell with the node representation from the previous time-step $\mathbf{h}_v^{i-1}$ and giving the incoming message in input to the GRU cell:
      \begin{equation}
      \label{eq:ggnn:3}
      \mathbf{h}_v^i = \text{GRU}(\mathbf{h}_v^{i-1}, \overline{\mathbf{a}_{v,v_0}^i,...,\mathbf{a}_{v,v_N}^i})~.
      \end{equation}
\end{enumerate}

Our final {\tt PUPPIML} model is pictorially represented in Fig.~\ref{fig:model}. The GGNN processing procedure is iterated stacking three GGNN layers one after the other. Each layer is characterized by its own node representation (100 for all of them), a set of $\mathbf{A}_t$ matrices and its own number of time steps (two for the first GGNN layer, one for the others). A given layer processes the output representation of the previous layer and creates a graph which is isomorphic to the input graph and characterized by a new vertex representation. In our case, the graph vertices are the particles in the event, and their input representation consists of the $(p_T,\eta,\phi)$ vector. Following Ref.~\cite{GGNN}, the initial input is added to the output of the next-to-last GGNN layer, with a residual connection analogous to what is done with ResNet~\cite{ResNet}, as shown in Fig.~\ref{fig:model}. In the last step of the network, the per-vertex (i.e., per-particle) output of the last GGNN layer is given as input to a fully connected DNN, which produces a pileup score, estimating the probability of a given particle to come from any of the pileupcollisions. The hyperparameters defining the {\tt PUPPIML} architecture are indicated in Fig.~\ref{fig:model}. We include as part of the graph the charged particles even when CHS is applied, as they aid in the classification of neighbouring particles. We train a different network for each mean pileup level (20, 80 and 140). Section~\ref{sec:robustness} discusses the generalization to a range of pileup configurations.

For our graph representation, we choose to connect all pairs of particles separated by a distance $\Delta R < R_1$ in the ($\eta$,$\phi$) plane, uniformly binning the distance $\Delta R$ into $N_0$ discrete edge types. 
We make all graphs undirected by introducing edges in both the forward and backward direction for each pair of connected particles. Moreover, as opposed to {\tt PUPPI}, we don't rescale the particles' four-momenta and simply discard all particles for which the network predicts a probability lower than a threshold $p_{\textrm{cut}}$. 
The choice of $R_1$, $N_0$, and $p_{\textrm{cut}}$ is made in order to maximise the network classification performance, as discussed in Sec.~\ref{sec:results}.

All our models are trained in Keras {\tt v2.1.2}~\cite{Keras} with a Tensorflow {\tt v1.2.1}~\cite{Tensorflow} backend, or in Tensorflow directly. The implementation of the GGNN is based on publicly available code~\cite{GGNNCode} published by Microsoft under the MIT license. We train the model using the Adam optimizer~\cite{Adam} with a learning rate of 0.004 and early stopping, and apply no pre-processing to our inputs. 

\section{Results}
\label{sec:results}

We make a general comparison of algorithm performances by comparing the ROC curves and the corresponding areas under the curve (auc) for the network models introduced in Section~\ref{sec:models} and the baseline rule-based approaches listed in Section~\ref{sec:intro}. For PUPPIML, we tested different architectures, varying the number of GGNN layers, the number of time steps, and the dimensionality of the node representation. We trained 20 randomly-picked configurations for these parameters and considered the one with best performance, measured in terms of the auc on the validation data. We observe a very weak dependence of the auc value on the network architecture. 

\begin{figure}[t!]
\centering %
\includegraphics[width=.5\textwidth]{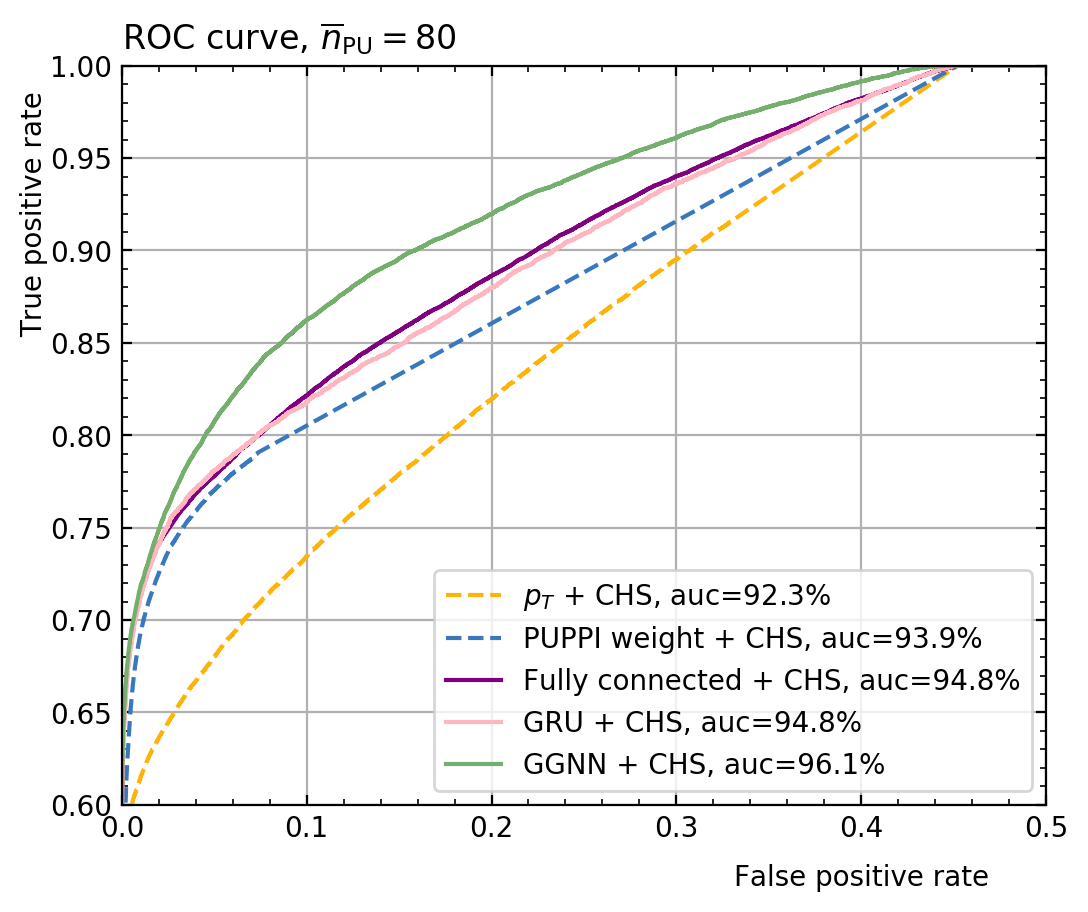}
\caption{\label{fig:1} Receiver Operating Characteristic (ROC) curve for our proposed features and models. Classifiers based on {\tt PUPPI} weight and $p_T$ are included as an indicator of the expected performance of {\tt PUPPI} and {\it SoftKiller} respectively. GGNN outperform other proposed architectures.}
\end{figure}

Using a sample of test data, we derive the auc values for different pileup configurations: a mean pileup $\overline{n}_{\textrm{PU}}= 20, 80$~or~$140$ when CHS is applied, and $\overline{n}_{\textrm{PU}}= 80$  when CHS is not used. 

\begin{table}[h!]
\centering
\begin{tabular}{rcccc}
\hline
$\overline{n}_{\textrm{PU}}$ & $20$ (CHS)      & $80$ (CHS)      & $140$ (CHS)     & $80$ (No CHS) \\ \hline
$p_T$ (SoftKiller)         & 92.3\%          & 92.3\%          & 92.5\%          & 64.9\%        \\
{\tt PUPPI} weight         & 94.1\%          & 93.9\%          & 94.4\%          & 65.1\%        \\
Fully-connected      & 95.0\%          & 94.8\%          & 94.8\%          & 68.5\%        \\
GRU                  & 94.8\%          & 94.8\%          & 94.7\%          & 68.8\%        \\ 
GGNN                 & \textbf{96.1\%} & \textbf{96.1\%} & \textbf{96.0\%} & \textbf{70.1\%}        \\ \hline
\end{tabular}
\caption{\label{tab:1} Area under the curve for the different discriminating variables and models. The highest values, highlighted in bold, are obtained when the GGNN architecture is used.}
\end{table}

As an example, Figure~\ref{fig:1} shows the ROC curves corresponding to $\overline{n}_{\textrm{PU}}= 80$ with CHS. Table~\ref{tab:1} reports the auc 
for this and the other pileupconfigurations. Since {\it SoftKiller} and {\tt PUPPI} base the decision of whether a particle comes from the LV on its transverse momentum or on the {\tt PUPPI} weight calculated from $\alpha^1$ respectively, the first two rows are meant as an indicator of their expected performances. For this purpose, we allow the {\tt PUPPI} weight to take on negative values if it is to the left of the median. We observe that the GGNN consistently obtains the best classification performance. Based on these results, we only consider the performance of our GGNN and the state of the art algorithms from here on. We tune $R_1 = 0.3$ and $N_0 = 5$ to maximize the auc. We fix the 
threshold parameter $p_{\textrm{cut}} = 0.4$ at $\overline{n}_{\textrm{PU}} = 20$ and $p_{\textrm{cut}} = 0.35$ at $\overline{n}_{\textrm{PU}} = 80$ or $\overline{n}_{\textrm{PU}} = 140$, so as to minimize the offset between the reconstructed and the LV observables. We find that minimizing the offset for one observable also approximately minimizes the offset for all other observables.

\begin{figure*}[th]
\centering 
\includegraphics[width=.95\textwidth]{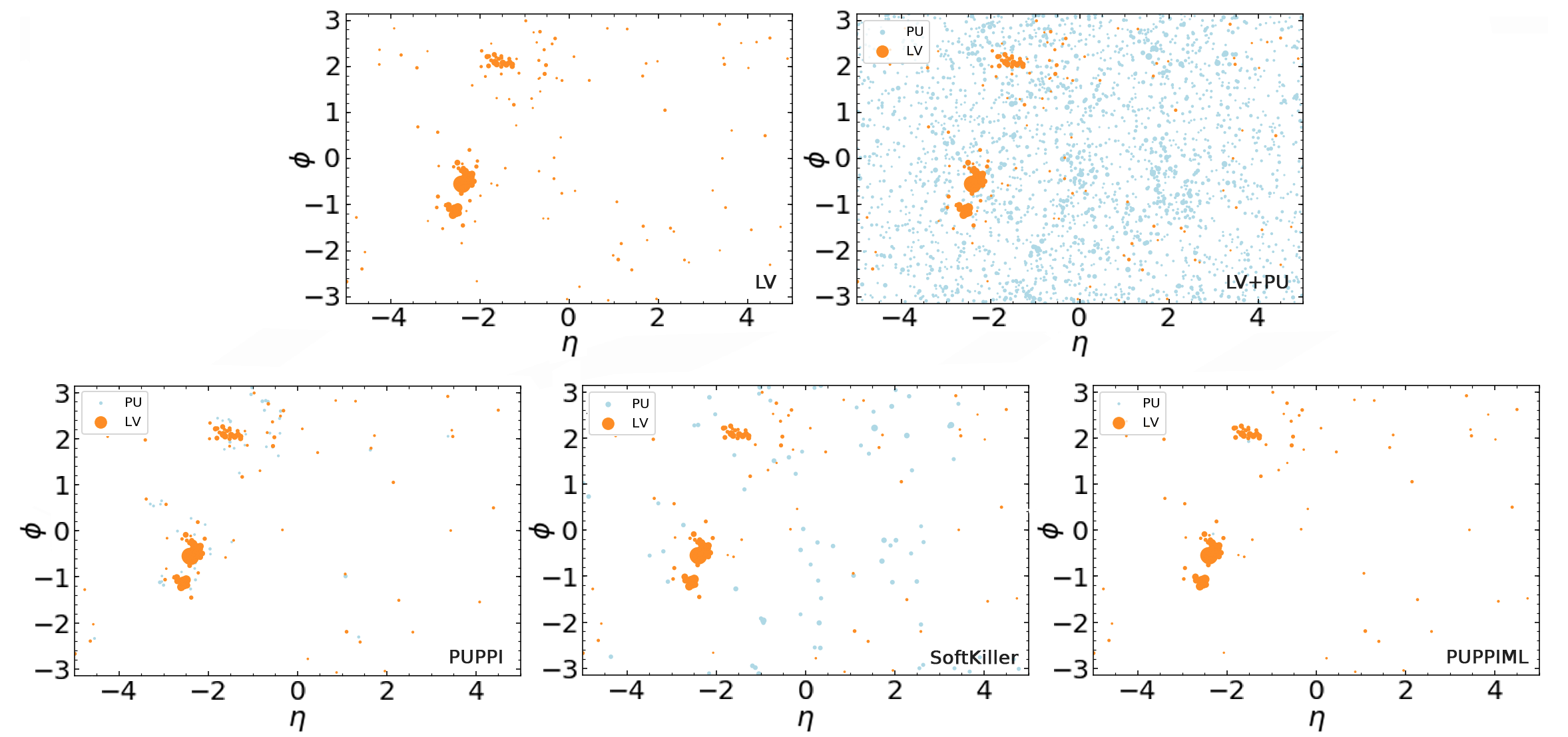}
\hfill
\caption{\label{fig:2} Depiction of the effect of running the different pileup mitigation algorithms on a random test event at $\overline{n}_{\textrm{PU}} = 20$. Particles from the LV are shown in orange (dark) and pileup particles are shown in blue (light). We show the ground truth (top left), the event contaminated by the pileup interactions (top right) and the reconstructed event after running {\tt PUPPI} (bottom left), {\it SoftKiller} (bottom center) and {\tt PUPPIML} (bottom right). {\tt PUPPIML} improves on {\tt PUPPI} by eliminating some of the low $p_T$ particles close to jets and on {\it SoftKiller} by eliminating the high $p_T$ particles that are far away from jets. All algorithms are run with CHS.}
\end{figure*}

\begin{figure*}[th]
\centering 
\includegraphics[width=.95\textwidth]{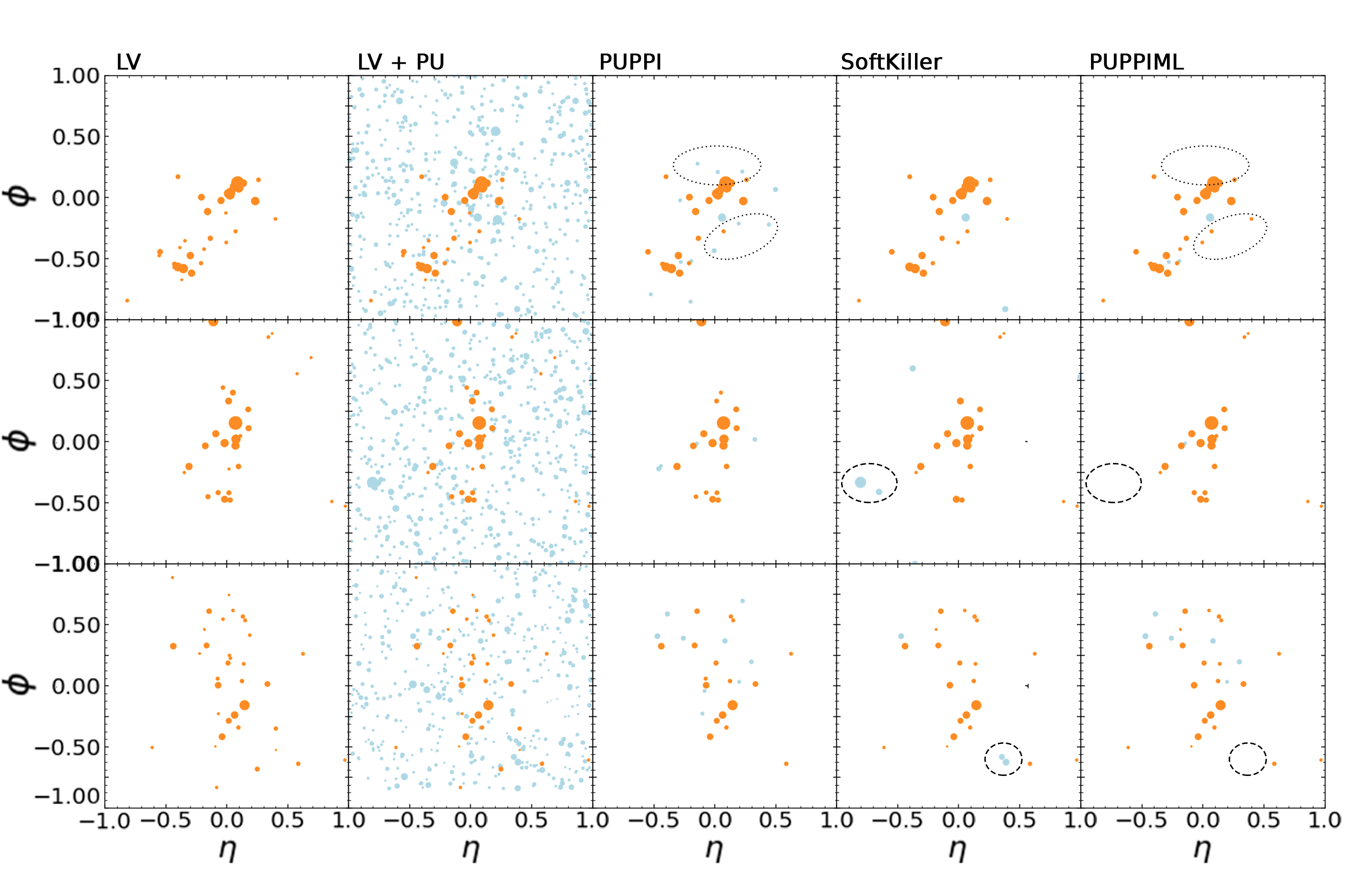}
\hfill
\caption{\label{fig:3} Depiction of the effect of running the different pileup mitigation algorithms on three jets at $\overline{n}_{\textrm{PU}} = 80$. Each circle represent one particle and the size of the circle is a function of the particle $p_T$. Particles from the LV are shown in orange (dark) and pileup particles are shown in blue (light). The top row shows the the ground truth, before (left) and after (right) adding the pileupcontribution. The bottom row shows the output returned by  {\tt PUPPI} (left), {\it SoftKiller} (center), and {\tt PUPPIML} (right), all running after applying CHS. {\tt PUPPIML} improves on {\tt PUPPI} by eliminating some of the low $p_T$ particles close to jets (dotted ellipses) and on {\it SoftKiller} by eliminating some of the high $p_T$ pileup particles that are far away from jets (dashed ellipses).}
\end{figure*}

\begin{figure*}[t]
\centering 
\includegraphics[width=.45\textwidth]{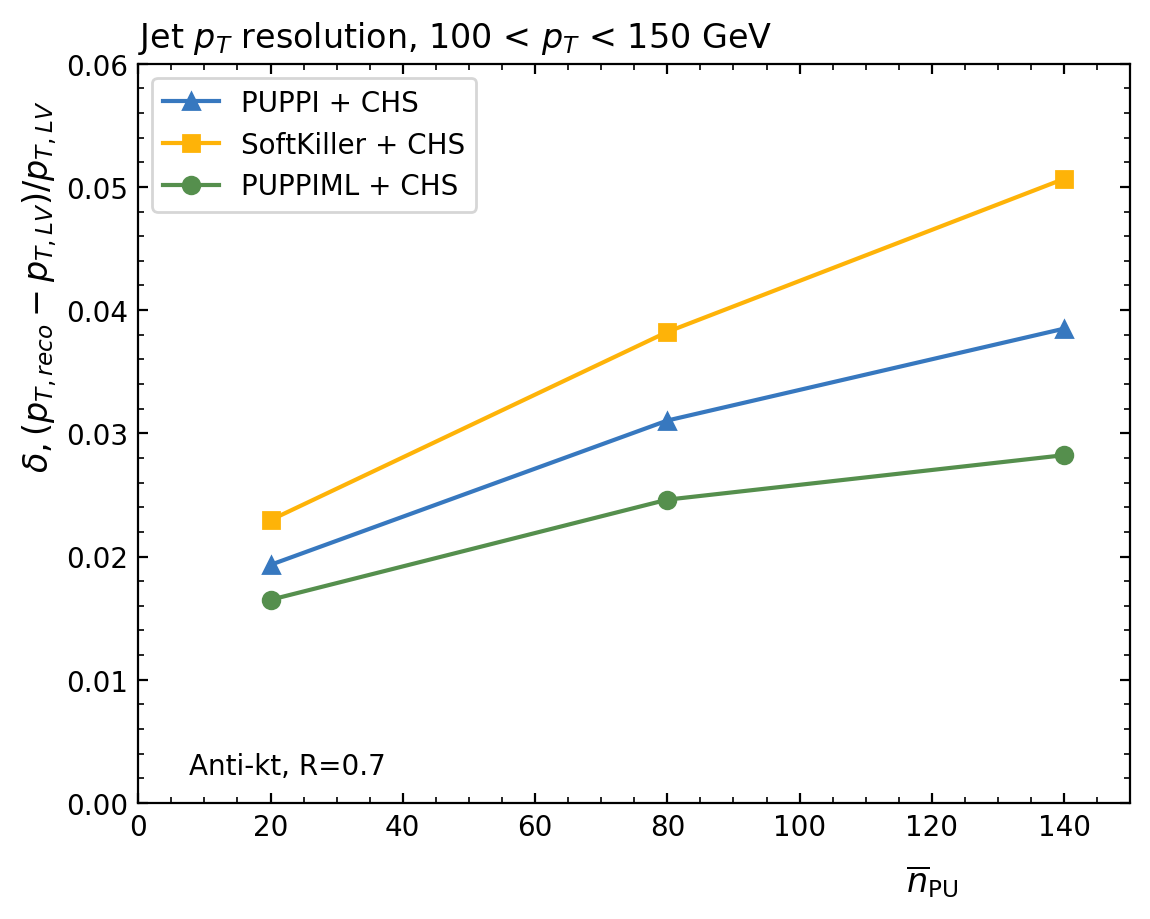}
\hfill
\includegraphics[width=.45\textwidth]{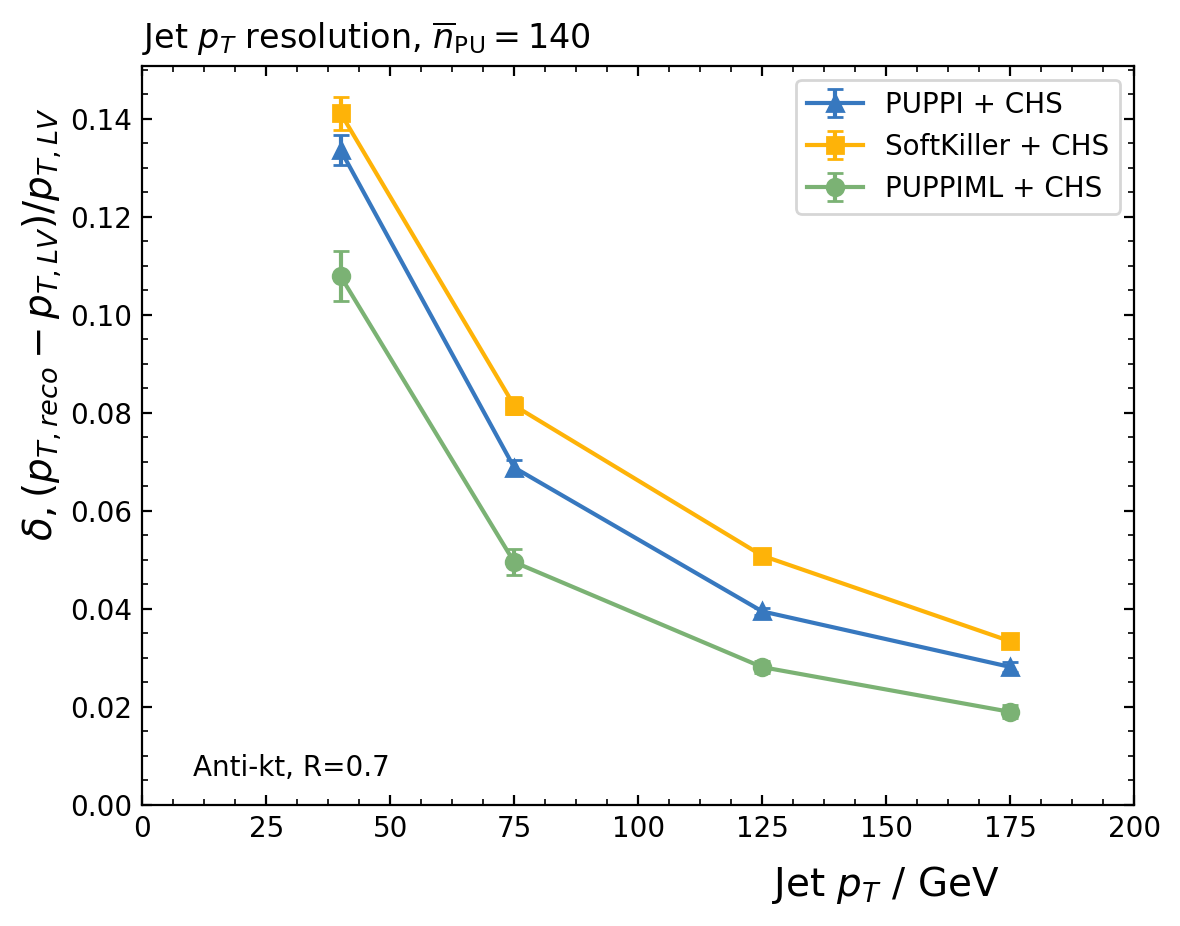}
\caption{\label{fig:4} Jet $p_T$ resolution as a function of $\overline{n}_{\textrm{PU}}$ for jets in the range 100 $< p_T <$ 150 GeV (top) and as a function of the jet transverse momentum at $\overline{n}_{\textrm{PU}} = 140$ (bottom) when CHS is applied.}
\end{figure*}

\begin{figure*}[t]
\centering 
\includegraphics[width=.45\textwidth]{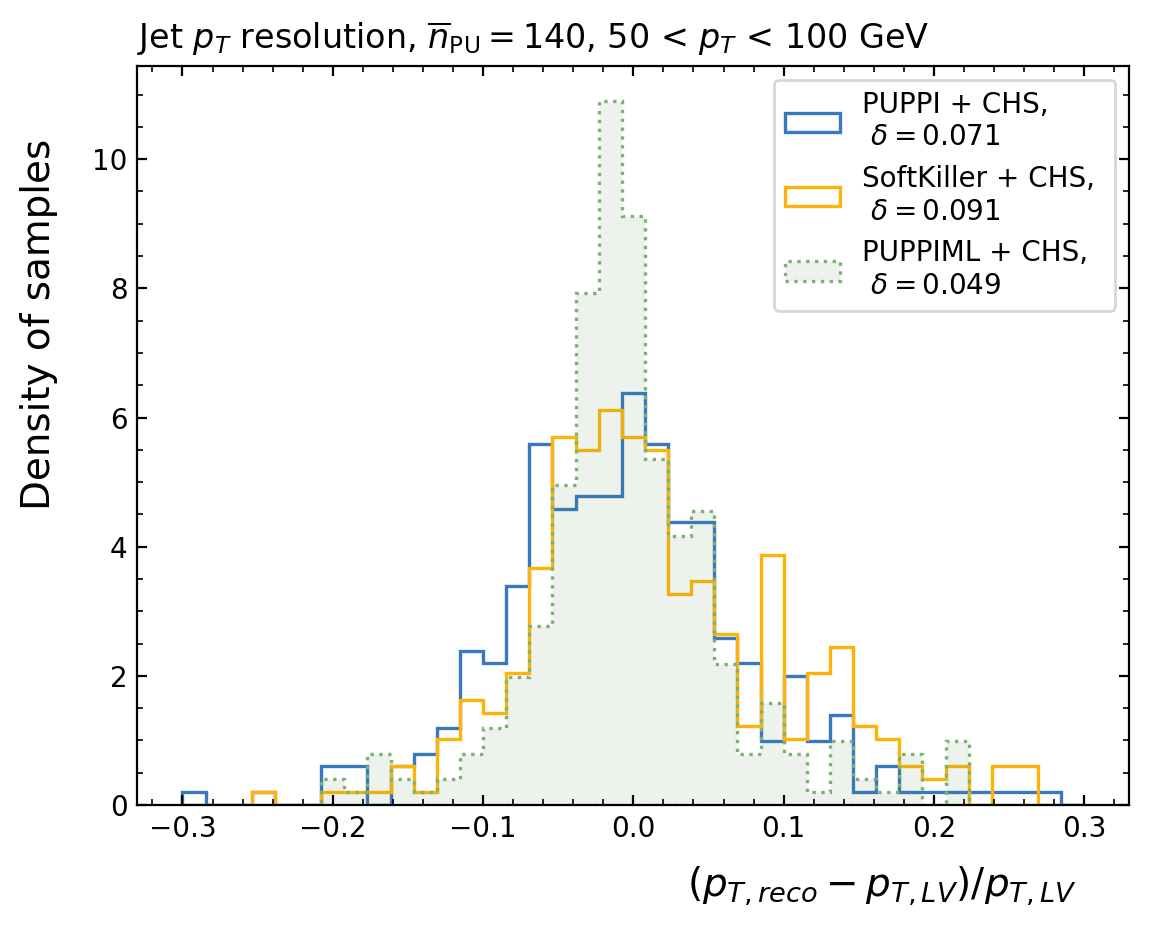}
\hfill
\includegraphics[width=.45\textwidth]{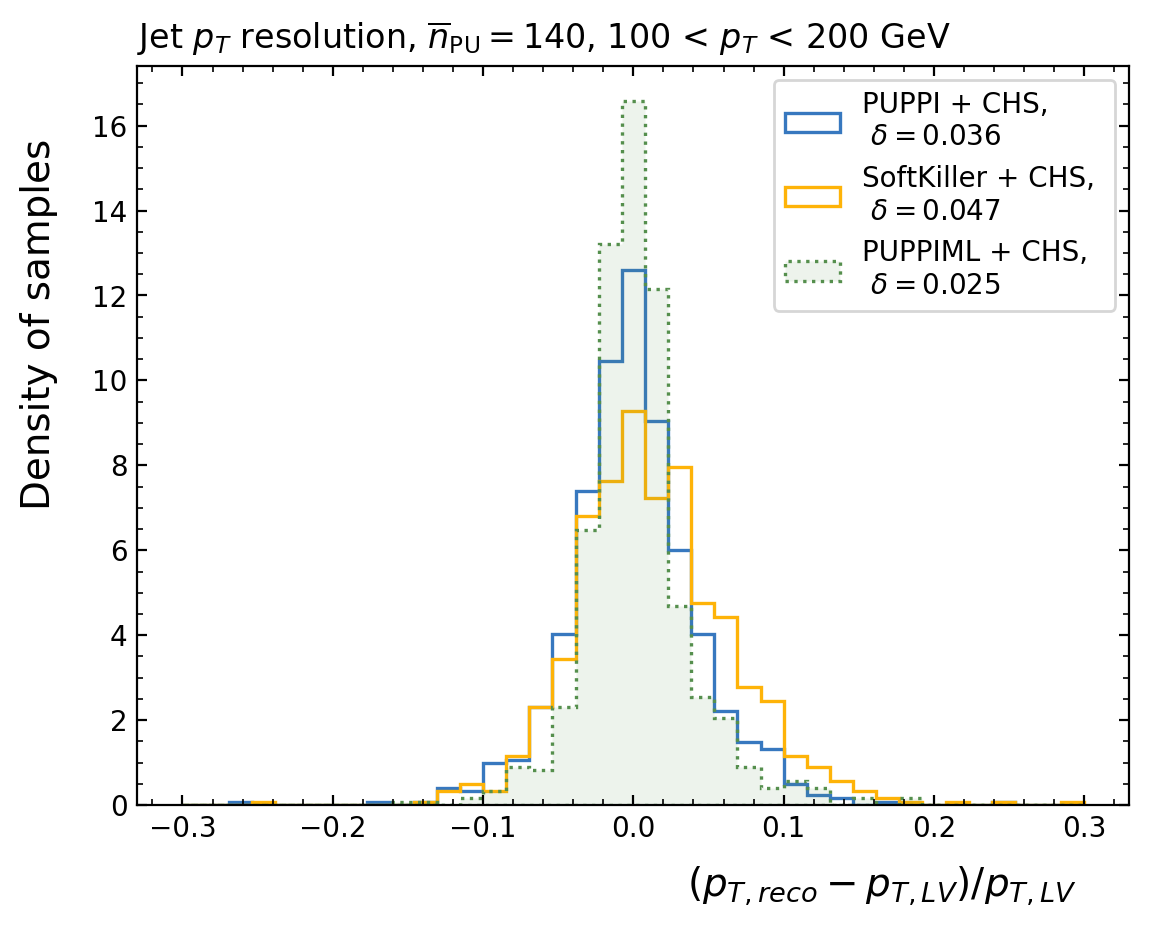}
\caption{\label{fig:5} Jet $p_T$ resolution at $\overline{n}_{\textrm{PU}} = 140$ for jets in the range 50 $< p_T <$ 100 GeV (top) and 100 $< p_T <$ 200 GeV (bottom) when CHS is applied. }
\end{figure*}

\begin{figure*}[t]
\centering 
\includegraphics[width=.45\textwidth]{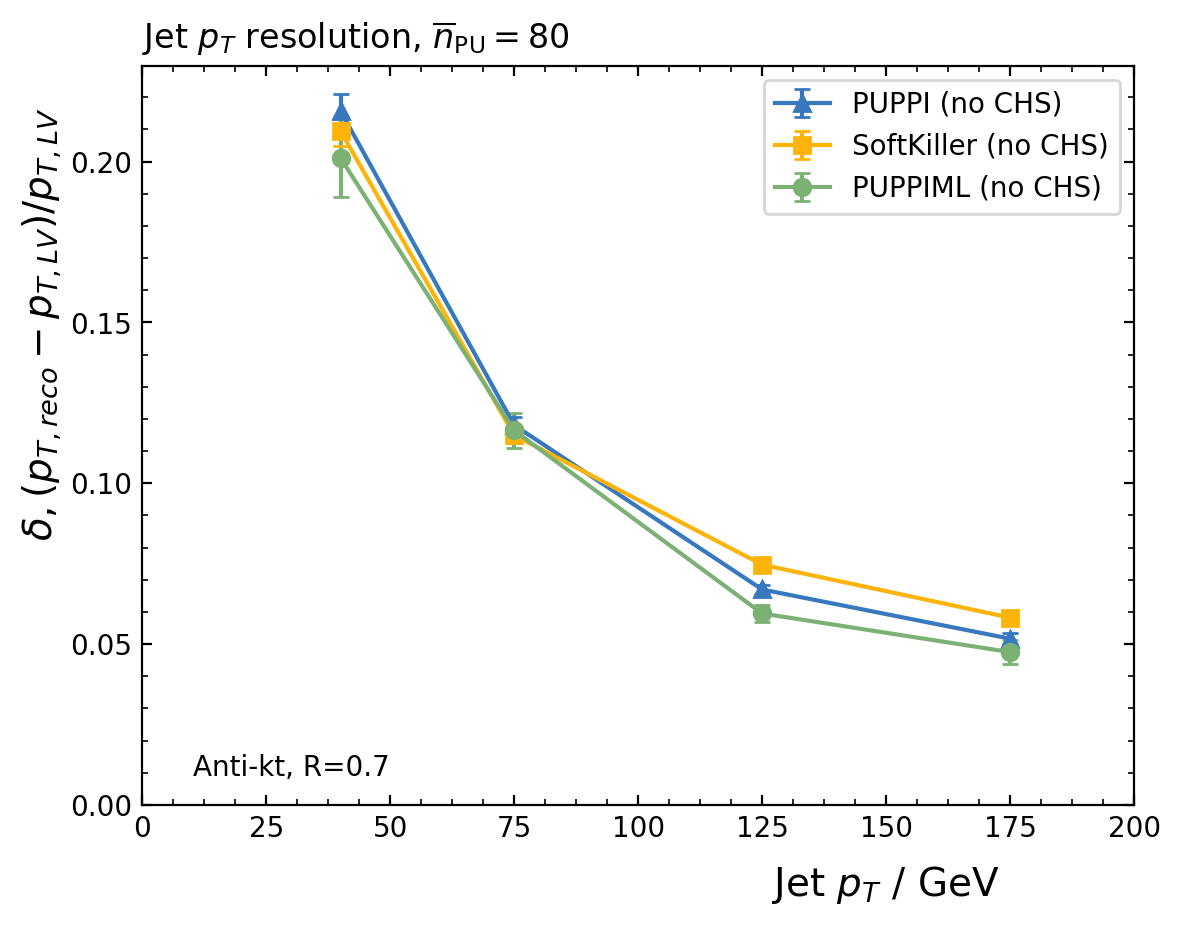}
\hfill
\includegraphics[width=.45\textwidth]{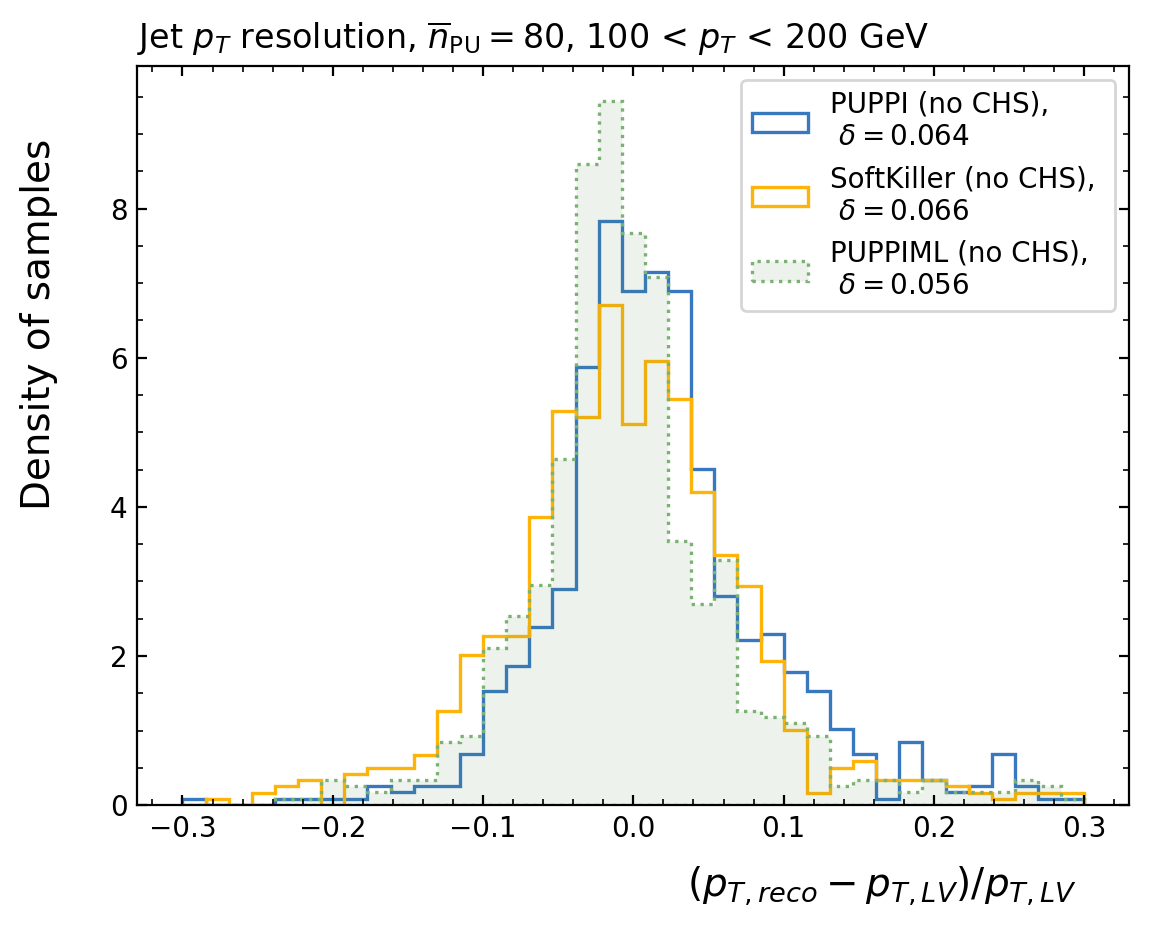}
\caption{\label{fig:noCHS} Jet $p_T$ resolution at $\overline{n}_{\textrm{PU}} = 80$ as a function of the jet $p_T$ (top) and for jets in the range 100 $< p_T <$ 200 GeV (bottom) when CHS is not applied.}
\end{figure*}

Figure~\ref{fig:2} shows the effect of running our proposed approach on an event at $\overline{n}_{\textrm{PU}}=20$. The event reconstructed by {\tt PUPPIML}  is shown on the bottom right, with particles represented as filled circles, sized according to their $p_T$. Dots are colored as orange (dark) if they come from the LV and blue (light) if they originate from pileup interactions. The event is also shown as reconstructed by {\tt PUPPI} (bottom left) and by {\it SoftKiller} (bottom center). Moreover, we show the ground truth on the top left and the unprocessed event on the top right. Using the same plotting conventions, Fig.~\ref{fig:3} shows the effect of the three PV mitigation algorithms for three jets in events with at $\overline{n}_{\textrm{PU}} = 80$. We note qualitatively that {\tt PUPPIML} improves on the state-of-the-art approaches, removing some low-$p_T$ pileup particles close to the jet that {\tt PUPPI} does not (dotted ellipses), and removing some high-$p_T$ particles far away from the jet that {\it SoftKiller} does not (dashed ellipses).

\subsection{Jet variables}

Following the methodology of Ref.~\cite{PUPPI}, we run the anti-k\textsubscript{t} clustering algorithm~\cite{antikt} using FastJet 3.3.0~\cite{fastjet1,fastjet2} with $R=0.7$ on both the LV-only and  full events, after applying {\tt PUPPI}, {\it SoftKiller}, and {\tt PUPPIML}. We consider a jet to match a LV-only one if the distance between the two in the  $(\eta, \phi)$ plane is less than $\Delta R_{\textrm{max}} = 0.3$. Figure~\ref{fig:4} shows the dependence of the jet $p_T$ resolution on $\overline{n}_{\textrm{PU}}$ and on the jet $p_T$. Figure~\ref{fig:5} shows this resolution for jets in the $50 < p_T < 100~\text{GeV}$ and the $100 < p_T < 200~\text{GeV}$ region for $\overline{n}_{\textrm{PU}} = 140$. We choose to quantify the resolutions of all indicated variables using the half-difference between the 14th and 86th percentile, $\delta = (P_{86} - P_{14})/2$, so as to avoid statistical fluctuations from potential outliers. Our method shows an improvement in resolution. Furthermore, such an improvement is larger for larger values of $\overline{n}_{\textrm{PU}}$, which makes the proposed solution particularly relevant for future LHC runs, when the average number of pileup interactions is expected to increase substantially. Mainly, at low pileup levels and very high $p_T$ the improvement with respect to the already adopted algorithms is marginal. We also test the resolution worsening when CHS is not applied. In this case, we train the network without the CHS flag in the dataset. Figure~\ref{fig:noCHS} shows that, while the resolution gain is smaller, there is still a small advantage in using {\tt PUPPIML}.

\begin{figure*}[htb]
\centering 
\includegraphics[width=.45\textwidth]{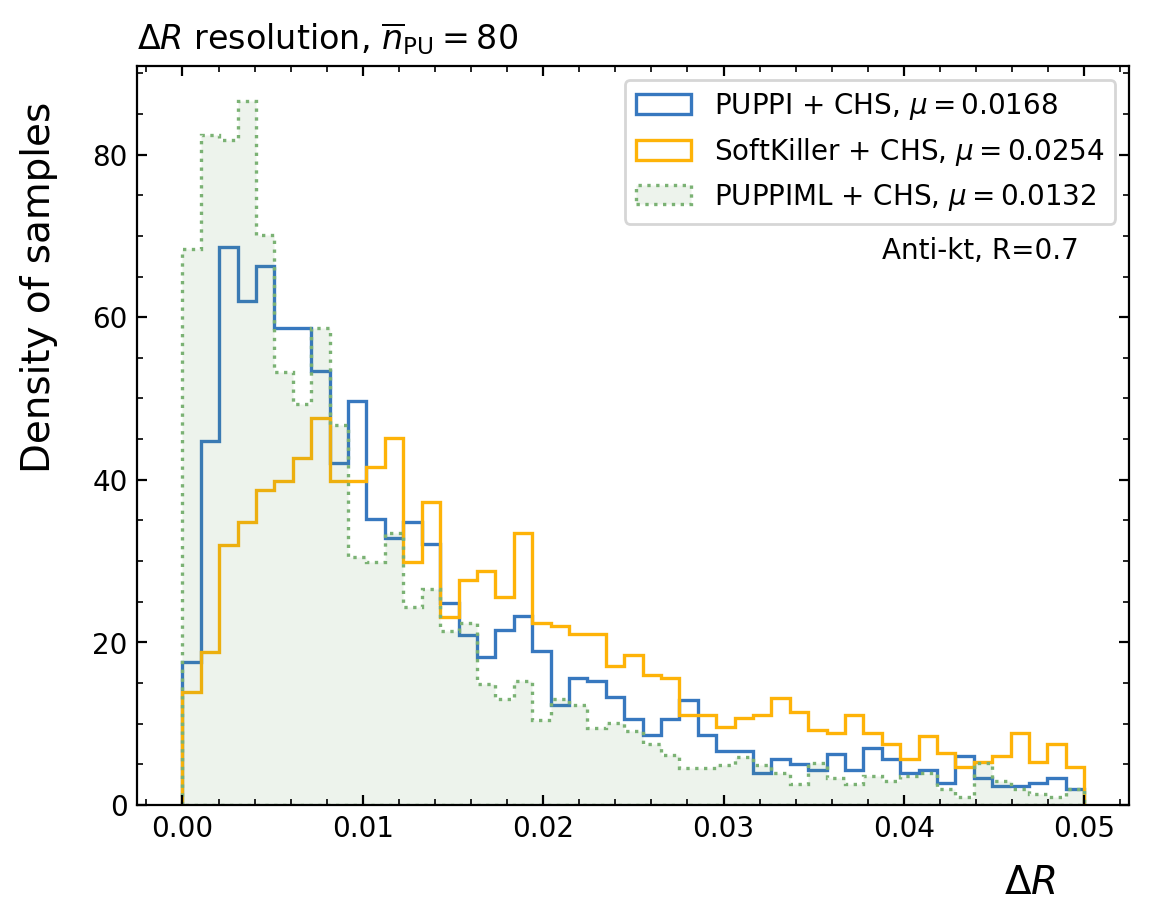}
\hfill
\includegraphics[width=.45\textwidth]{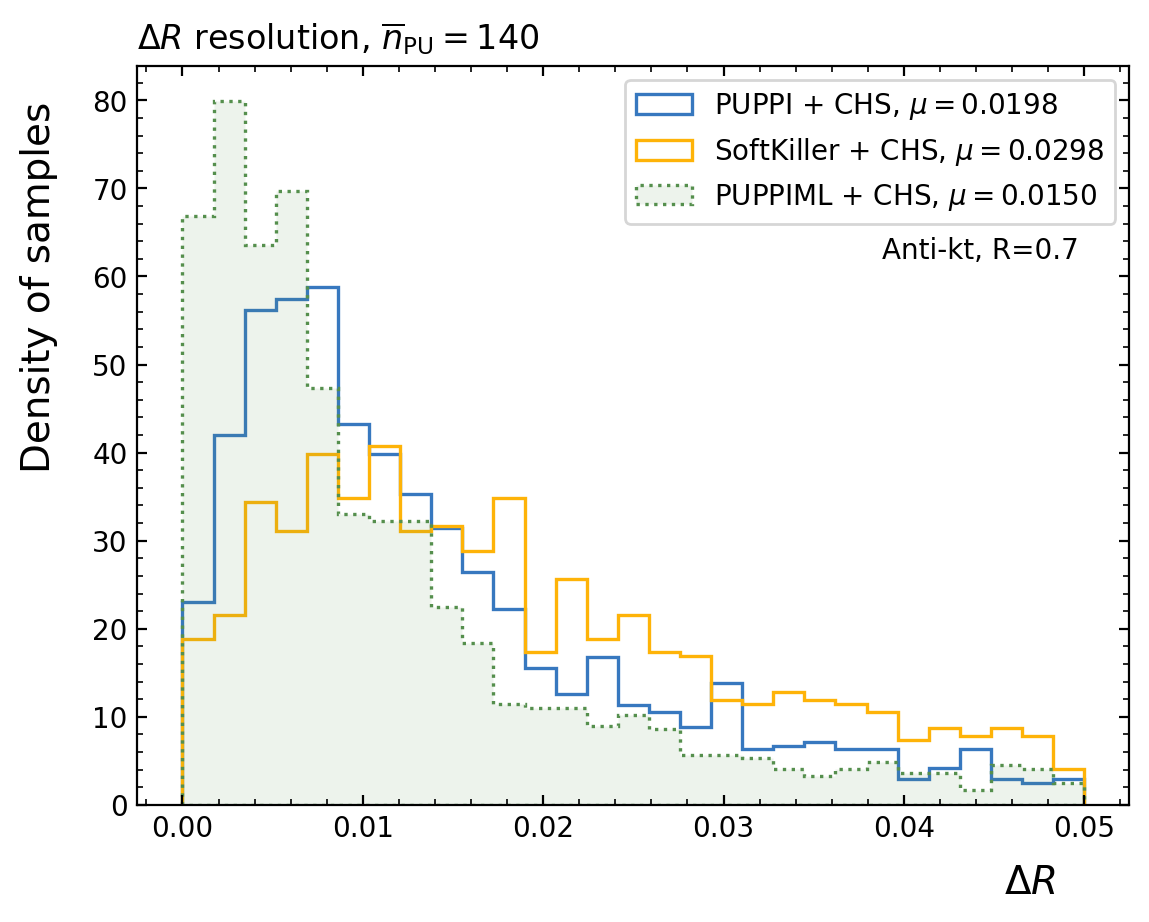}
\caption{\label{fig:6} Distribution of $\Delta R = \sqrt{\Delta \eta^2 + \Delta \phi^2}$ between the jet in absence and in presence of pileup effects, for $\overline{n}_{\textrm{PU}} = 80$ (top) and $\overline{n}_{\textrm{PU}} = 140$ (bottom). CHS is applied to all algorithms.}
\end{figure*}

Figure~\ref{fig:6} shows how the mitigation algorithms affect the resolution of the jets position in the $(\eta, \phi)$ space, quantified through the mean $\Delta R$. {\tt PUPPIML} shows an improvement of $\sim 15\%$ in resolution at $\overline{n}_{\textrm{PU}}=20$, $\sim 25\%$ at $\overline{n}_{\textrm{PU}}=80$, and $\sim 30\%$ at $\overline{n}_{\textrm{PU}}=140$. 

\begin{figure*}[htb]
\centering 
\includegraphics[width=.45\textwidth]{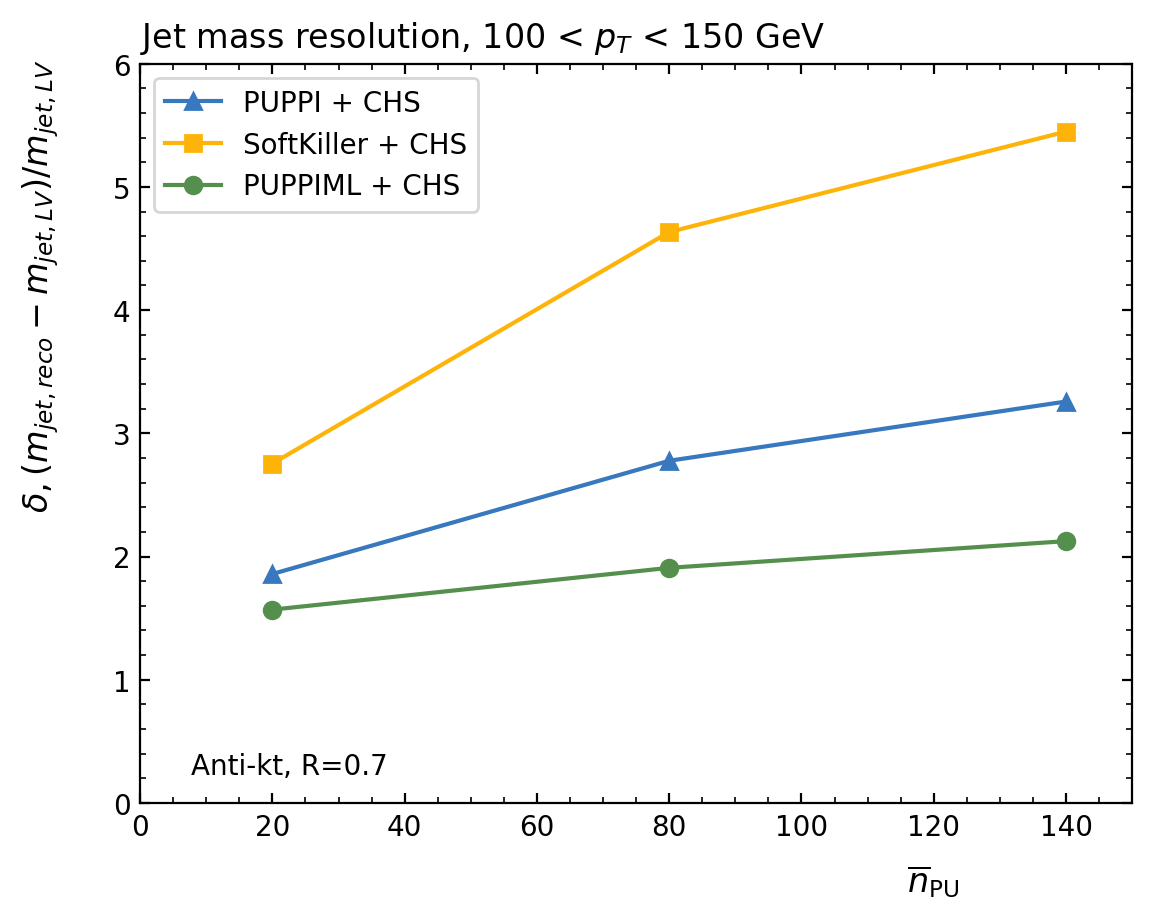}
\hfill
\includegraphics[width=.45\textwidth]{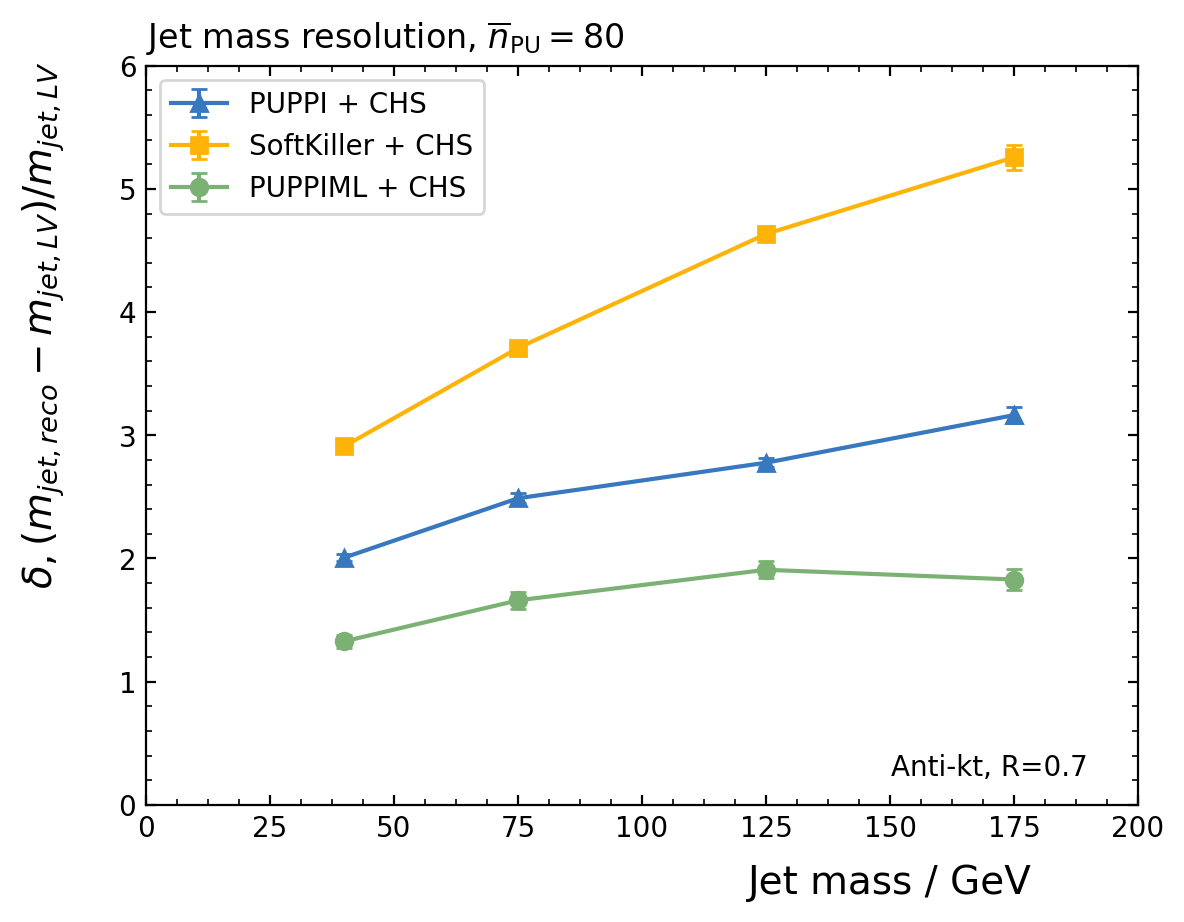}
\caption{\label{fig:7}  Jet mass resolution as a function of $\overline{n}_{\textrm{PU}}$ for jets in the range 100 $< p_T <$ 150 GeV (top) and as a function of the jet transverse momentum at $\overline{n}_{\textrm{PU}} = 80$ (bottom) when CHS is applied.}
\end{figure*}

Finally, in order to study the reconstruction of jet shapes, we consider the jet-mass resolution, which is commonly taken as a proxy to evaluate a pileup-removal performance. Figure~\ref{fig:7} shows the resolution as a function of $\overline{n}_{\textrm{PU}}$ and of the jet $p_T$. We find that a better mass resolution can be obtained when {\tt PUPPIML} is applied. 

\begin{figure*}[htb]
\centering 
\includegraphics[width=.45\textwidth]{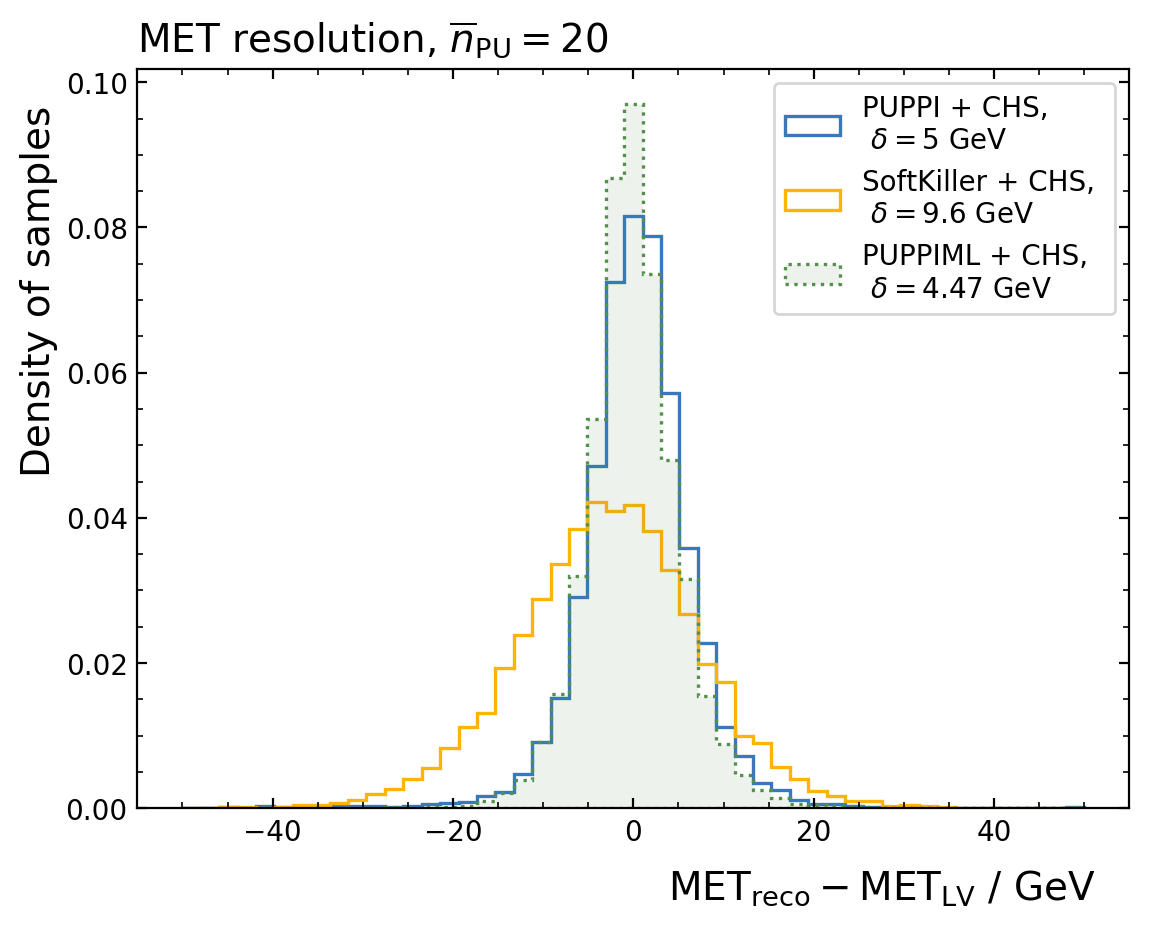}
\hfill
\includegraphics[width=.45\textwidth]{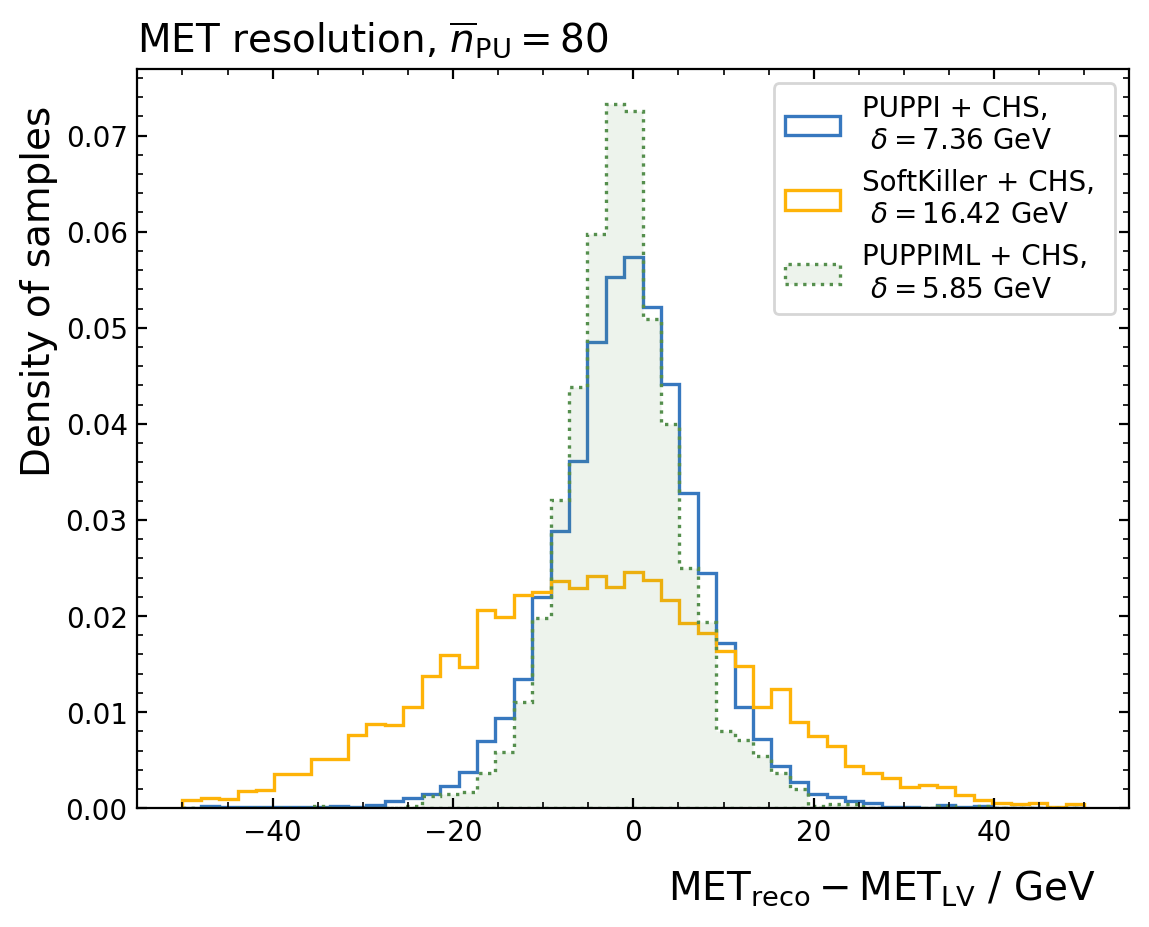}
\caption{\label{fig:8} Missing transverse energy resolution at $\overline{n}_{\textrm{PU}} = 20$ (top) and $\overline{n}_{\textrm{PU}} = 80$ (bottom). One should notice that, due to the sample definition, the events are characterized by real MET, corresponding to the $Z$ $p_T$.}
\end{figure*}

Since our network takes the full event as input, we expect it to be able to correct not only jet-related quantities but also global ones. Figure~\ref{fig:8} shows that this is indeed the case, when the missing transverse energy (MET) in the event is considered. Since this variable involves summing over all present particles, its sensitivity to pileup contamination is typically stronger. 

\subsection{Robustness}
\label{sec:robustness}

It is important that the network learns a pileup mitigation strategy that is effective across samples and not overly dependent on the specific features of the training dataset. We probe this dependence by evaluating the network auc performance on (i) different mean pileup levels, (ii) different {\tt PYTHIA8} tunes and (iii) different decay processes. Table~\ref{tab:npu} shows that the performance degrades when evaluating the network at highly different pileup distributions. Nevertheless, a model trained at multiple pileup levels is effective over a wide range. Tables~\ref{tab:tune} and~\ref{tab:processes} show that {\tt PUPPIML} reaches auc values larger than the corresponding {\it SoftKiller} and {\tt PUPPI} ones, even when using different tunes and on unseen decay channels respectively. This feature is particularly important in view of the fact that pileup mitigation algorithms are designed and tuned on simulation to then be used on data.

\begin{table}[th!]
\centering
\begin{tabular}{crcccc}
\multicolumn{1}{l}{}              & \multicolumn{1}{l}{}          & \multicolumn{4}{c}{Trained on  $\overline{n}_{\textrm{PU}}$}                                    \\ \cline{2-6} 
\multirow{4}{*}{\smash{\rotatebox[origin=c]{90}{{\vtop{\hbox{\strut Evaluated}\hbox{\strut on  $\overline{n}_{\textrm{PU}}$}}}}}}             & \multicolumn{1}{l}{\textbf{}} & \textbf{20}  & \textbf{80}  & \textbf{140} & \textbf{20+80+180}    \\ \cline{2-6} 
                                 & \textbf{20}                & \textbf{96.1\%} & 95.6\%          & 95.0\%          & \textbf{96.1\%} \\
                                  & \textbf{80}                & 95.7\%          & \textbf{96.1\%} & 95.9\%          & \textbf{96.1\%} \\
                                  & \textbf{140}               & 95.1\%          & 96.0\%          & \textbf{96.0\%} & \textbf{96.1\%} \\ \cline{2-6} 
\end{tabular}
\caption{\label{tab:npu} Area under the curve for {\tt PUPPIML}, trained and evaluated on different pileup configurations. We train models at $\overline{n}_{\textrm{PU}}$ equal to 20, 80 and 140, and on a dataset containing equal splits of the three. While the performance decreases when testing on previously unseen pileup configurations, it is still superior to the {\it SoftKiller} and {\tt PUPPI} proxies (see Table~\ref{tab:1}). A model trained on the combined dataset is capable of learning a strategy that generalizes to a wide range of pileup levels. Some results are highlighted for visual clarity.}
\end{table}

\begin{table}[!ht]
\centering
\begin{tabular}{rccc}
\hline
\multicolumn{1}{l}{} & \textbf{Tune 4C} & \textbf{Tune 1} & \textbf{CUEP8M1} \\ \hline
$p_T$  (SoftKiller)        & 92.5\%           & 91.7\%          & 92.4\%           \\
{\tt PUPPI} weight         & 94.4\%           & 93.7\%          & 94.2\%           \\
\textbf{\tt PUPPIML}     & \textbf{96.0\%}  & \textbf{95.4\%} & \textbf{95.9\%}  \\ \hline
\end{tabular}
\caption{\label{tab:tune} Area under the curve (auc) for different pileupmitigation algorithms. The pileup configuration is fixed to $\overline{n}_{\textrm{PU}} = 140$. The models are trained on $Z \rightarrow \nu \overline{\nu}$ events, generated with {\tt PYTHIA} 4C tune and then tested on the same process, generated with different {\tt PYTHIA8} tunes. For {\tt PUPPIML}, the most significant drop of performance is observed using Tune 1, yet with an auc value larger than the corresponding {\it SoftKiller} and {\tt PUPPI} values, also shown for comparison.}
\end{table}

\begin{table}[!ht]
\centering
\begin{tabular}{rccc}
\hline
\multicolumn{1}{l}{} & \boldmath{$Z \rightarrow \nu \overline{\nu}$} & \boldmath{$H \rightarrow b \overline{b}$} & \boldmath{$H \rightarrow g g$} \\ \hline
$p_T$   (SoftKiller)       & 92.5\%          & 92.0\%          & 92.0\%           \\
{\tt PUPPI} weight         & 94.4\%          & 94.0\%          & 94.0\%           \\
\textbf{\tt PUPPIML}     & \textbf{96.0\%} & \textbf{95.7\%} & \textbf{95.7\%}  \\ \hline
\end{tabular}
\caption{\label{tab:processes} Area under the curve (auc) for different pileupmitigation algorithms. 
The pileup configuration is fixed to $\overline{n}_{\textrm{PU}} = 140$. The models are trained on $Z \rightarrow \nu \overline{\nu}$ events, generated with {\tt PYTHIA} 4C tune and then tested on different processes, generated with the same {\tt PYTHIA} tune. The {\tt PUPPIML}  model is able to generalize to decays involving multiple jets and including bottom quarks and gluons with only a small decrease in auc performance. The corresponding values for the {\it SoftKiller} and {\tt PUPPI} proxies are shown for comparison.}
\end{table}

\subsection{Computational Performance Considerations}

We run all our experiments on a Titan Xp GPU and observe speeds of $\mathcal{O}(100 \text{ events/s})$ at $\overline{n}_{\textrm{PU}}=20$, $\mathcal{O}(20 \text{ events/s})$ at $\overline{n}_{\textrm{PU}}=80$ and $\mathcal{O}(10 \text{ events/s})$ at $\overline{n}_{\textrm{PU}}=140$ for the network forward propagation. This is more than an order of magnitude slower than {\it SoftKiller} and {\tt PUPPI}. We note, however, that our implementation is not tuned for computational performance. We verified that reducing the number of layers, the number of edges and size of the hidden representation can result in improvements of a factor of 2 with a reduction in auc of less than $0.2\%$. Moreover, the network requires a pre-processing step in which the neighbours to each particle are found; while this is slow if done through pairwise brute force (as we did in our study, trading simplicity for computation efficiency), the computational cost can be reduced substantially if implemented through a heap or other more advanced data structures. Similarly to {\tt PUPPI}, removing a large portion of the pileup particles (e.g., by applying a threshold on the particles' $p_T$) results in much faster jet clustering downstream, partially compensating for the time spent to perform the pileup mitigation.

We further note that our approach maintains locality, since each particle is connected only to those particles with distance $\Delta R < R_1$ from it. By choosing $R_1=0.3$ and with a total of four time-steps across the three GGNN layers, only particles local neighborhood of $\Delta R < 1.2$ can influence the classification of a given particle. Taking advantage of this local nature, one could further speed up the {\tt PUPPIML} inference by parallelizing it. Asynchronous Gated Graph Neural Networks and "towers" in Message Passing Neural Networks are a step in this direction and can handle overlapping parts of the graph.

\section{Conclusions and outlook}
\label{sec:conclusions}

This paper introduces a new pileup mitigation algorithm, {\tt PUPPIML}, that extends the commonly-used {\tt PUPPI} algorithm through Deep Learning via Graph Neural Networks. The proposed algorithm shows resolution improvements in all the considered jet- and event-related quantities. For jet-related quantities, {\tt PUPPIML} provides a resolution improvement up to $\sim 30\%$ with respect to {\tt PUPPI} and even more with respect to {\tt SoftKiller}. 

The algorithm is designed to return the probability that a given particle belongs to the leading vertex of the event (as opposed to originating from pileup). Its output can be used to weight each particle by its probability to originate from the leading vertex. On the other hand, it can also be used for pileup removal by tuning a threshold on the probability value returned by the classifier. When tuned on different pileup configurations, this threshold is found to be mildly dependent on the details of the training sample. 

By representing an event as a graph of reconstructed particles, {\tt PUPPIML} allows to avoid any assumptions on the geometry or granularity of a given detector. In this respect, it represents an interesting alternative to the approach presented in Ref.~\cite{PUMML}. 

In terms of future work, we believe that a further exploration of similar locality-preserving architectures, in particular variations of Message Passing Neural Networks, could further improve performances. Adversarial training, requiring the network to correct the full event directly as opposed to simply classifying individual particles,   could aid in improving the resolution gain on jet substructure quantities.

\section*{Acknowledgments}
We are grateful to  Caltech and the Kavli Foundation for their support of undergraduate student research in cross-cutting areas of machine learning and domain sciences. This work was conducted at  ``\textit{iBanks}", the AI GPU cluster at Caltech. We acknowledge NVIDIA, SuperMicro  and the Kavli Foundation for their support of ``\textit{iBanks}".
This project has received funding from the European Research Council (ERC) under the European Union's Horizon 2020 research and innovation program (grant agreement n$^o$ 772369). This project is partially supported by the United States Department of Energy, Office of High Energy Physics Research under Caltech Contract No. DE-SC0011925.

\bibliographystyle{epj}
\bibliography{bib}

\end{document}